\documentclass{aa}
\usepackage{graphics}

\newcommand{\ls}    {LS 992/RX J0812.4--3114}
\def\simless{\mathbin{\lower 3pt\hbox
     {$\rlap{\raise 5pt\hbox{$\char'074$}}\mathchar"7218$}}}   
\def\simmore{\mathbin{\lower 3pt\hbox
     {$\rlap{\raise 5pt\hbox{$\char'076$}}\mathchar"7218$}}}   

\begin{document}

\thesaurus{06 
              (08.09.2;  
               08.02.3;  
	       08.16.6;  
               08.05.2);  
	       13.25.5;  
	       }

\title{The Be/X-ray binary LS 992/RX J0812.4--3114: physical parameters
and long-term variability\thanks{Based on observations collected at the
South African Astronomical  Observatory, the European Southern
Observatory, Chile (ESO N64.H-0059) and the Teide Observatory, Tenerife
(Spain)}}

\subtitle{}

\author
{P. Reig \inst{1,2}, I. Negueruela \inst{3}, D.A.H. Buckley\inst{4}, 
M.J. Coe \inst{5}, J. Fabregat\inst{6}, N.J. Haigh\inst{5}}

\institute{
Foundation for Research and Technology-Hellas. GR-711 10 Heraklion. Crete. 
Greece. 
\and Physics Department. University of Crete. GR-710 03 Heraklion. Crete.
Greece.
\and SAX SDC, Agenzia Spaziale Italiana, c/o Telespazio, via Corcolle 19, 
I-00131 Roma, Italy.
\and South African Astronomical Observatory, PO Box 9, Observatory 7935. 
Cape Town. South Africa
\and Physics \& Astronomy Department, Southampton University. SO17 1BJ. UK.
\and Astronomy \& Astrophysics Department. University of Valencia. 
E-46100 Burjassot-Valencia. Spain.
}

\authorrunning{Reig et al.}
\titlerunning{The Be/X-ray binary LS 992/RX J0812.4--3114}

\offprints{pablo@physics.uoc.gr}

\date{Accepted \\
Received : 2 August 2000\\
}

\maketitle

\begin{abstract} 

We present the first long-term optical and infrared study of the
optical counterpart to the source RX J0812.4--3114, an X-ray pulsar with a
Be type companion. During the period covered by the observations the
profile of some Balmer lines changed from absorption to emission and back
again to absorption. Contemporaneously, the infrared magnitudes varied by
more than 0.8 mag. This long-term variability is interpreted as the
formation and subsequent dissipation of the Be star's disc. The building
up of the disc ended up in an active X-ray state characterised by regular
outbursts occurring at 80 day intervals. The overall duration of the
formation/dissipation of the disc is found to be $\simless 4.3 $ years. 
Optical spectroscopic and infrared photometric observations were used to
refine the spectral type of the primary (B0.2IV) and to monitor the
circumstellar  envelope around the Be star. UBVRI and $uvby\beta$
photometric observations allowed the determination of the astrophysical
parameters of the optical companion.

\end{abstract}

\keywords{stars: individual:  -- LS 992/RX J0812.4--3114
                binaries: general -- 	
                stars: pulsars: general --
		stars: emission-line, Be --
		X-rays: stars 
		}

\section{Introduction}

The star LS 992 is the optical counterpart to the X-ray source RX
J0812.4--3114. Such an association was originally proposed by Motch et al.
(1997), as a result of a systematic cross-correlation between the ROSAT
All Sky Survey (Voges et al. 1999) and several OB star catalogues in the
SIMBAD database.  The X-ray light curve of \ls\ is characterised by
31.88 second pulsations, while the X-ray spectrum is best represented by
an absorbed power-law component with a exponentially cut-off (Reig \&
Roche 1999). In December 1997 the source made a transition from a
quiescent state to a flaring state (Corbet \& Peele 2000), in which
regular flares separated by 80 day intervals were detected with the
All-Sky Monitor (ASM) onboard the {\em Rossi X-ray Timing Explorer}.
Corbet \& Peele (2000) attributed the origin of these flares to the
periastron passage of the neutron star, hence this periodicity was
naturally associated with the orbital period. The peak X-ray luminosity
during outbursts was estimated to be of the order of 2.3 $\times$
10$^{36}$ erg s$^{-1}$, assuming a distance to the source of 9 kpc.

In the optical band, LS~992 appears as a relatively bright $V\sim$~12.4
B0--1III-V star (Reed 1990). The optical spectra revealed H$\alpha$ in
emission and evidence for short-term ($\sim$ weeks) V/R variability (Motch
et al. 1997).  The combination of an emission line B star plus a neutron
star, whose existence is inferred from the characteristics of the X-ray
emission, defines a Be/X-ray binary. Be/X-ray  binaries  comprise 
approximately  70\% of the more  general class of high mass X-ray 
binaries,  the remaining  $\sim$  30\%  contain evolved  (luminosity  class
I and II) primaries. The high energy radiation is the result of accretion
of the material  expelled by the Be star onto the neutron star.

In this work we present the first long-term study of the optical
counterpart of RX J0812.4--3114 and discuss the consequences of the
correlated behaviour in the optical, infrared and X-ray wavelength bands.

\section{Observations}

Blue spectroscopy of the source was obtained on 31st October 1999 using
the 1.52-m telescope at La Silla Observatory, Chile. The telescope was
equipped with the Boller \& Chivens spectrograph + \#32 holographic 
grating and the Loral 38 camera. This configuration gives a resolution of
$\sim 0.5$ \AA/pixel. Measurements of arc line FWHM  indicate a spectral
resolution of $\approx 1.4$ \AA\ at $\sim 4500$ \AA. A second spectrum
was obtained on 15th September 2000, this time using holographic
grating \#33, which gives a resolution of
$\sim 1.0$ \AA/pixel (FWHM$\approx 3.0$ \AA\ at $\sim 4500$ \AA).

All other spectra were taken using the 1.9m telescope at SAAO, South Africa
equipped with the Cassegrain spectrograph. The 1996 data were obtained using
the Reticon Photon Counting System (RPCS), whereas all subsequent data were
obtained using the {\it SITe2} CCD detector. Details of the wavelength
coverage achieved in each data set are given in Table 1. 

The optical photometric data were acquired at the SAAO 1.0-m telescope 
using the Tek8 CCD camera, employing $UBVR_CI_C$, $uvby\beta$ and
H$\alpha$ filters, at a plate scale of 0.35 arcsec $\rm pixel^{-1}$ for
the January 1999 observations and 0.7 arcsec $\rm pixel^{-1}$ for the
October 1996 observations. The CCD frames were  pre-processed using IRAF's
CCDPROC package for bias subtraction, overscan  removal and flat-fielding.
Aperture photometry was performed in the STARLINK  GAIA package, using
observations of E-region standard stars (Menzies et al.  1989) for
calibration to the Stromgren system.

The infrared photometric data were obtained at two telescopes - the 1.9m 
equipped with the MkIII IR photometer at SAAO, in South Africa and the
1.5m Telescopio Carlos  Sanchez (TCS), Tenerife, Spain using the
Continuously Variable Filter (CVF).

\begin{table}
\begin{center}
\caption{Journal of the optical spectroscopic observations}  
\label{sp}
\vspace{4mm}
\begin{tabular}{lcccc}
\hline
Date	  & MJD		&Wavelength 	  &EW(H$\alpha$)&EW(H$\beta$) \\
	  &		&coverage (\AA)   &(\AA)	&(\AA) \\
\hline
\multicolumn{5}{c}{low-resolution spectra} \\
\hline
07-10-96  &50364	&3200--8000	&--6.5$\pm$1.0   &--0.6$\pm0.4$   \\
05-02-98  &50850	&3800--7800	&--16.8$\pm$0.8  &--1.4$\pm$0.5  \\
\hline
\multicolumn{5}{c}{red-end spectra}\\
\hline
04-04-96  &50178	&6300--7000	& +1.4$\pm$0.5  &\\
03-02-98  &50848	&6300--7000	&--17.0$\pm$1.0 &\\
09-01-99  &51188	&6200--6900	&--20.5$\pm$0.8 &\\
\hline
\multicolumn{5}{c}{blue-end spectra}\\
\hline
05-04-96  &50179        &4200--5000     & & +1.6$\pm$0.4 \\
05-03-98  &50878	&3700--5200	& &--1.4$\pm$0.3  \\
10-01-99  &51189	&3700--5500	& &--2.2$\pm$0.3  \\
31-10-99  &51483	&3800--5000	& &--1.8$\pm$0.1  \\
15-09-00  &51803	&3000--5300	& & +2.1$\pm$0.1  \\
\hline

\end{tabular}
\end{center}
\end{table}

\begin{table*}
\begin{center}
\caption{Journal of photometric observations}  
\label{phot}
\vspace{4mm}
\begin{tabular}{lcccccccccccc}
\hline
\multicolumn{13}{c}{optical observations} \\
\hline
Date   &   MJD  &Telescope &I    &Ierr   & R    &Rerr   & V  &Verr   & B
&Berr & U	&Uerr\\
05-10-1996 & 50362   &SAAO1.0m   & 11.72  & 0.02  & 12.16  & 0.02  & 12.48   &
0.02  & 12.86  & 0.02  & 12.57  & 0.09 \\
\hline
Date   &   MJD  &Telescope &y    &yerr   & b    &berr   & v  &verr   & u
&uerr &$\beta$ & $\beta$err\\
23-01-1999 & 51202  &SAAO1.0m & 12.52 & 0.04 & 12.89 & 0.04 & 13.11
&0.07 &13.33 & 0.14 &2.47 & 0.02\\
\hline
\multicolumn{13}{c}{infrared observations} \\
\hline
Date   &   MJD  &Telescope &J    &Jerr   & H    &Herr   & K  &Kerr   & L' &L'err &\\
13-01-1996 & 50096.56  &TCS  & 11.39 & 0.05 & 11.14 & 0.02 & 10.84 & 0.02 &    & &\\
06-02-1998 & 50851.40  &SAAO1.9m & 11.21 & 0.01 & 10.95 & 0.01 & 10.70 & 0.02 & 10.36 & 0.21 &\\
07-02-1998 & 50852.51  &SAAO1.9m & 11.25 & 0.01 & 11.01 & 0.01 & 10.66 & 0.07 &    &  &\\
08-02-1998 & 50853.43  &SAAO1.9m & 11.26 & 0.02 & 11.02 & 0.01 & 10.77 & 0.02 & 10.61 & 0.63 &\\
28-10-1998 & 51115.76  &TCS  & 11.67 & 0.03 & 11.31 & 0.02 & 11.03 & 0.04 &    &  & \\
05-01-1999 & 51184.53  &SAAO1.9m & 11.54 & 0.02 & 11.25 & 0.07 & 10.95 & 0.02 &    &  &\\
22-01-2000 & 51566.59  &TCS  & 12.01 & 0.08 & 11.89 & 0.04 & 11.43 & 0.01 &    &  &\\
22-01-2000 & 51566.593 &TCS  & 12.04 & 0.33 & 11.78 & 0.14 & 11.25 & 0.02 &    &  &\\
\hline
\end{tabular}
\end{center}
\end{table*}
\begin{figure*}
\mbox{}
\vspace{8.0cm}
\includegraphics{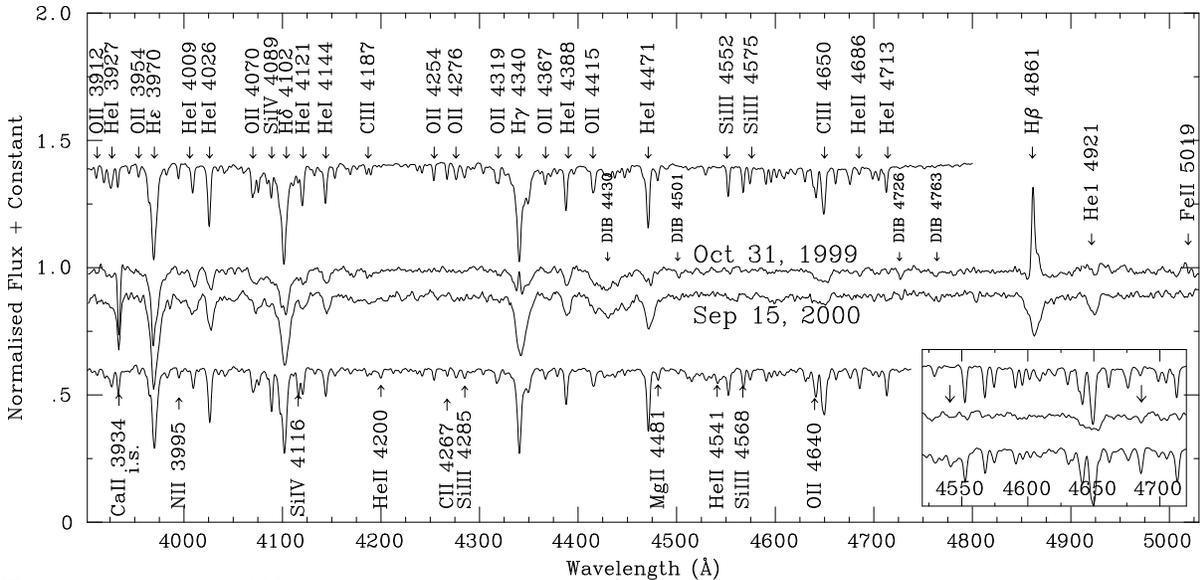}
\caption[]{The spectrum of LS~992 in the classification region (middle) is 
compared 
to those of the MK standard stars HD~48434 (B0III, bottom) and 1~Cas 
(B0.5III, top). The inset shows the smooth progression of 
\ion{He}{ii}~$\lambda\lambda$ 4541, 4686 \AA\ (marked with arrows) when
compared to the neighbouring \ion{Si}{iii}~$\lambda 4553$ \AA\
and \ion{O}{ii}~$\lambda 4676$ \AA. All spectra have been smoothed with
a $\sigma=0.8$ Gaussian filter, normalised by division into a spline fit
to the continuum and offset for display.} 
\label{class}
\end{figure*}

\section{Spectral type}

The spectra of LS~992 in the classification region are displayed in
Fig.~\ref{class}. The September 2000 spectrum shows no evidence for
emission components, but it has a lower resolution and Signal-to-Noise
Ratio (SNR) than the October 1999 spectrum. It has been included, however,
so that the obvious changes can be observed. In particular, we call
attention to the very strong in-filling of the \ion{He}{i} lines at
$\lambda\lambda$ 4026, 4471, 4921 \AA\, typical of Be/X-ray binary
counterparts. In the October 1999 spectrum H$\beta$ is strongly in
emission, with an asymmetric blue-dominated profile, while similarly
asymmetric emission components  are present in the cores of all Balmer
lines down to H$\epsilon$.  Longwards of $\lambda \sim 4750$ \AA\ the
continuum is dominated by weak (presumably \ion{Fe}{ii}) emission
features.

Photospheric lines are all shallow and broad indicating
a fast rotator. However, given the high SNR of the spectrum even the
weakest metallic lines are detected. As is typical of Be/X-ray binary
counterparts, the presence of weak \ion{He}{ii} lines indicates a spectral
class close to B0. In order to improve the classification we have compared
the spectrum of the source with those of standard stars taken either at
the same resolution with the same instrument or at higher resolution
(from the list of Steele et al.1999).

We have used the main sequence standards $\nu$ Ori (B0V),  $\tau$ Sco
(B0.2V) and $o$ Per (B0.5V). The absence of \ion{He}{ii}~$\lambda 4200$
\AA\ at this resolution indicates that the star is later than B0. If the
star was in the  main sequence, it would indicate a spectral type B0.5V or
later.  However, the strength of \ion{He}{ii}~$\lambda 4686$ \AA\ when
compared to the nearby \ion{O}{ii}~$\lambda 4676$ \AA\ feature and the
detectability of \ion{He}{ii}~$\lambda 4541$ \AA\ argue against this later
classification. This seems to indicate a higher luminosity class.

In Fig.~\ref{class}, the spectrum of LS~992 is shown together with those
of the MK standard stars HD~48434 (B0III) and 1~Cas (B0.5III). The smooth 
progression of all the \ion{He}{ii} lines favours a spectral
classification B0.2III for LS~992. However, some metallic lines seem
weaker than expected and a luminosity class IV sounds more appropriate.
Direct comparison with the spectrum of the B0.2IV standard $\phi^{1}$ Ori
at a slightly lower resolution (from the digital atlas of Walborn \&
Fitzpatrick 1990) supports this interpretation. Therefore we will accept a
spectral type B0.2IV, though B0.2III cannot be excluded.

\section{Astrophysical parameters}

The determination of the astrophysical parameters by means of photometric
calibrations in a Be star is not as straightforward as in a non-emission
line B-type star due to the presence of the surrounding envelope, which
distorts the characteristic photometric spectrum. Thus  one has to correct
for both circumstellar and interstellar reddening before any  calibration
can be applied. We have made use of the iterative procedure developed by
Fabregat \& Torrej\'on (1998, hereafter FT98) to determine the intrinsic
magnitudes. This method is based on the fact that there exists a
correlation between the equivalent widths of the Balmer lines and the
anomalies in the $uvby$ photometric indices produced by continuum
circumstellar emission. 

\subsection{Extinction, absolute magnitude and distance}
\label{red}

By applying the FT98 method we are able to disentangle the relative
contributions of the interstellar and circumstellar extinction: 
$E^{\rm{is}}(b-y)=0.433\pm0.033$ and $E^{\rm{cs}}(b-y)=0.051\pm0.025$,
respectively. The errors reflect the accuracy of the FT98 procedure. Using
the relation $E^{\rm{is}}(B-V)=1.35\, E^{\rm{is}}(b-y)$ for interstellar
extinction (Crawford \& Mandwewala 1976) and $E^{\rm{cs}}(B-V) = 1.20 \,
E^{\rm{cs}}(b-y)$ for circumstellar extinction (Fabregat \& Reglero 1990)
we find a {\em total} value for the extinction of $E(B-V)=0.65\pm0.05$,
which is in good agreement with the values of 0.69 (from optical
photometry) and 0.8$\pm$0.2 (from interstellar spectral lines) reported 
by Motch et al. (1997). Correcting the $B-V$ index for extinction we find
$-0.27$, consistent with a B0.2IV star (Wegner 1994).

Although the $\beta$ index (Crawford \& Mander 1966) provides a measure of
the luminosity for O and B stars it is strongly affected by emission. Thus
it cannot be used  to calibrate the stellar luminosity. In view of the
good quality of our spectrum we prefer to assume the luminosity class
derived from the spectral analysis and take $M_V=-4.3\pm0.5$ as would
correspond for a B0.2 giant-subgiant star. The error has been arbitrarily
taken so that it includes disagreements between different $M_V$
calibrations. The distance, estimated from the distance-modulus relation
is found to be $d\approx 8.8$ kpc.

\subsection{Effective temperature}

A B0.2IV star has a temperature close to 30000 K (Panagia 1973, Lamers
1981, Vacca et al. 1996). In O and B type stars the index $c_0$ is closely
related to the effective temperature (Crawford 1978).  Using the
calibration $\log\, T_{\rm{eff}}=0.186c_0^2-0.580c_0+4.402$ (Reig et al. 1997a)
we obtain $T_{\rm{eff}}=25400\pm4500$ K, where the error was obtained by
propagating the photometric error. Due to the uncertainty in the $u$
measurement, hence $c_0$, the value of the temperature is not well
constrained. Also, possible residual circumstellar
emission may result in slightly larger $c_0$ values.
If the intrinsic colour derived above, $(B-V)_0=-0.27$, is used in
combination with the calibration of Gulati et al. (1989) a 
value of $26400\pm2300$ K is found. Whereas if Eq.(11) of Napiwotzki
et al. (1993) is applied, $T_{\rm{eff}}\approx29500$ K.
As a compromise, we adopted a temperature of $28000\pm2000$ K

Directly related to the effective temperature there is  the bolometric
correction, $BC=M_{\rm{bol}}-M_V$. The calibrations of $BC$ in terms of
$T_{\rm{eff}}$ result in $BC=-2.8\pm0.2$ (Balona 1994, Vacca et al. 1996).

\begin{table}
\begin{center}
\caption{Astrophysical parameters of LS 992} \label{param}
\begin{tabular}{ll}
Spectral type   &B0.2IV--III  \\
E(B-V)          &0.65$\pm$0.05  \\
$T_{\rm{eff}}$       &28000$\pm$2000 K  \\
Radius          &10$\pm$2 $R_{\odot}$  \\
Mass            &17$\pm$3 $M_{\odot}$  \\
$M_V$           &--4.3$\pm$0.5\\
Distance        &8.8$\pm$4.0 kpc  \\
BC              &--2.8$\pm$0.2\\
$M_{\rm{bol}}$  &--7.1$\pm$0.5\\
$\log g$        &3.7$\pm$0.2 cm$^2$ s$^{-1}$\\
$v\, \sin i$    &240$\pm$20 km s$^{-1}$\\
\end{tabular}
\end{center}
\end{table}

\subsection{Radius}

In order to estimate the radius of LS 992 we have re-written the
luminosity equation $L=4\pi \sigma R^2 T^4_{\rm{eff}}$ in terms of the
visual brightness parameter $F_V$ (Barnes \& Evans 1976) and the $(b-y)_0$
colour index. With the value of the reddening derived in Sect.~\ref{red}
we obtain $(b-y)_0=-0.115$ which gives $F_V=4.166\pm0.026$ according to
Eq. (6) of Moon (1984), where the error is that of the least-squares fit
of the calibration. Substituting in $F_V=4.236-0.1
M_V-0.5\log(R/R_{\odot})$ (Moon 1985) we finally obtain $R=10.0\pm2.6$
R$_{\odot}$. The final error was obtained by propagating the errors
through  the luminosity equation. This value compares well to the average
radii in terms of MK spectral types calculated by Moon (1985), who give 12
R$_{\odot}$ and 7.6 R$_{\odot}$ for B0III and B1III stars, respectively.
And also to those of Vacca et al. (1996) who derived 8 and 14.8
R$_{\odot}$ for B0.5V and B0.5III stars, respectively.

\subsection{Mass and Gravity}

With the  values of the temperature  and  bolometric luminosity, we can
estimate the mass from evolutionary models. We have used the interpolation
formula given by Balona (1994) for the Claret \& Gim\'enez (1992) models
and found M = 17$\pm$3~M$_{\odot}$. Once we know the mass and radius of
the star we can calculate its gravity by means of $\log g=4.44+\log
M-2\log R$, where $M$ and $R$ must be given in solar units. The result is
$\log g=3.7\pm 0.2$ cm s$^{-1}$.

\subsection{Rotational velocity}

The projected rotational velocity $v\, \sin i$ can be estimated from the
width of the spectral lines.  However, in practice, it is difficult to
find clean spectral lines, i.e., not affected by circumstellar emission.
The spectrum taken in September 2000 shows Balmer lines in absorption,
indicating a disc-loss phase. In addition, \ion{He}{I} lines are mainly
formed in the star's photosphere and are expected to have the smallest
disc contribution. We have used the \ion{He}{I} $\lambda$4026, \ion{He}{I}
$\lambda$4144 and \ion{He}{I} $\lambda$4388 lines and obtained FWHM$= 6.1,
5.9$ and 7.0 \AA, respectively.

From Buscombe's (1969) approximation

\begin{equation}
\frac{v\, \sin i}{c}=\frac{FWHM}{2\lambda_0 (\ln2)^{1/2}}
\end{equation}

\noindent where $\lambda_0$ is the rest wavelength of the line and $FWHM$
represents the full width at half maximum corrected for instrumental
broadening, we obtain $v\, \sin i=$240 km s$^{-1}$, 220 km s$^{-1}$ and 260
km s$^{-1}$ for \ion{He}{I} $\lambda$4026, $\lambda$4144 and 
$\lambda$4388 respectively. Similar values are found using the fits of
Steele et al. (1999).

\begin{figure}
\mbox{}
\vspace{10cm}
\includegraphics{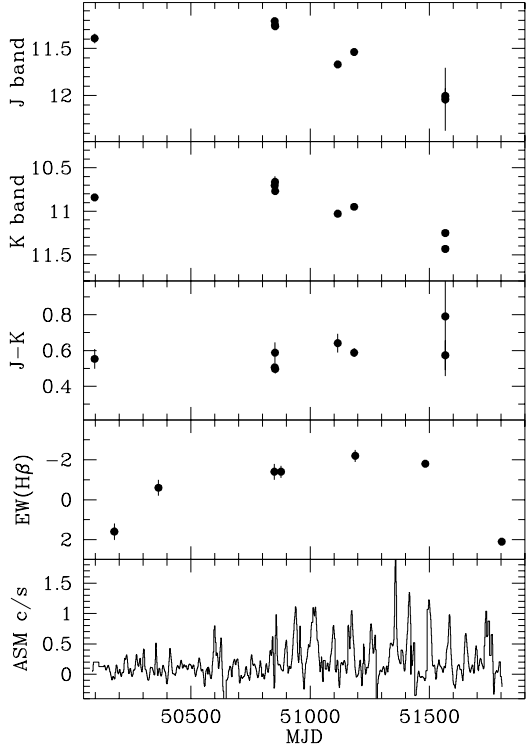}
\caption{IR magnitudes, H$\beta$ equivalent width and ASM light curve for
the period April 1996--September 2000. The activity of the source increased
suddenly at around 1997 December (MJD 50800)}
\label{long-term}
\end{figure}
\begin{figure}
\mbox{}
\vspace{10cm}
\includegraphics{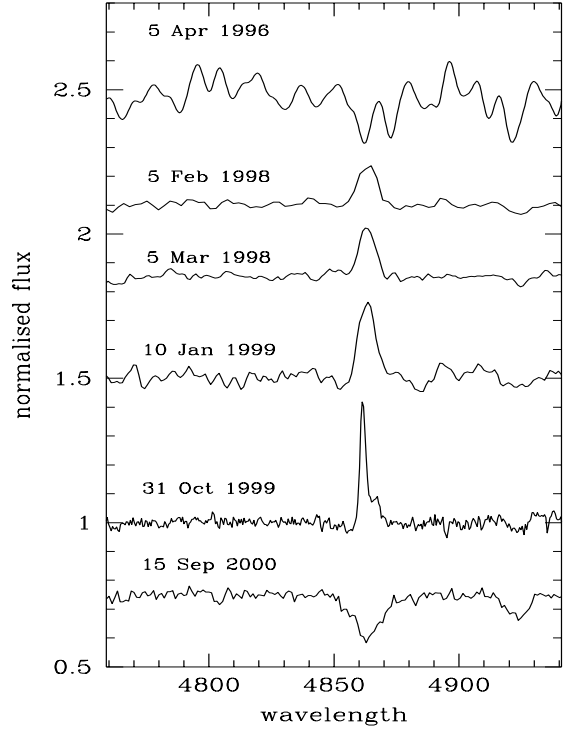}
\caption{The large line profile variability indicates major structural changes 
in the disc of the Be star. The continuum has been normalised to unity and the 
spectra offset for display.}
\label{beta}
\end{figure}
\section{Long-term variability}

Prior to 1998 \ls\ was in an X-ray dormant state as indicated by the fact
that the  ASM flux was consistent with the X-ray background flux, $\sim$
0.1 c/s (Fig~\ref{long-term}). At the end of 1997 (MJD $\sim$ 50800) the
source underwent a transition from this inactive state to a flaring state,
consisting of regular outbursts occurring at approximately 80 day intervals
(Corbet \& Peele 2000). At the onset of the X-ray activity the optical
companion was in a bright infrared and optical state.

In Be/X-ray binary systems the optical primary has a circumstellar disc,
which is presumably made up of material expelled by the B star. A neutron
star orbits around the Be star in a moderately eccentric orbit. The
ultimate physical process which leads to mass ejection in Be stars is
unknown but it is probably one (or a combination) of these: radiation
pressure, centrifugal force, thermal expansion and hydromagnetic waves
(Strafella et al. 1998; Balona 2000). This circumstellar disc is
responsible for the two defining observational properties of a Be star,
namely the infrared excess and the emission lines. The mechanism that
powers the X-ray source is accretion onto a neutron star and the fuel is
the material in the disc. 

In this scenario we would expect some type of correlation between the
X-ray and the optical/infrared emission. Fig~\ref{long-term} shows the ASM
and infrared light curves together with the evolution of the H$\beta$
equivalent width. When the circumstellar disc becomes large enough the
neutron star is able to interact with it and produce X-rays. As can be
seen in Fig~\ref{long-term} the equivalent width of the H$\beta$ line,
EW(H$\beta$) increased slowly during the years prior to the onset of the
X-ray activity, indicating the building-up of the disc. Spectral hydrogen
lines were in absorption in April 1996, in emission in October 1996 and
back to absorption in Semptember 2000.  The absorption profile indicates
that the Be star's disc was at some point non existent. Subsequently, the
disc began to grow until it was large enough for the neutron star to
interact. The source was X-ray active in December 1997. Thus, the build-up
of the disc took $\simless$ 18 months. The source reached an optical and
infrared  maximum and then began to decrease. 

The issue of how far/big the disc can grow was discussed by Reig et al.
(1997b), who concluded that the disc may be truncated by the continuous
passage of the neutron star once the disc radius is comparable to the
periastron distance. This effect is expected to be stronger in systems
with short orbital periods. Assuming that the 80 day periodicity found in
the X-rays (Corbet \& Peele 2000) represents the orbital period, the
diagram EW(H$\alpha$)--P$_{\it{orb}}$ (Reig et al. 1997b) predicts
EW$_{\it{max}}$ $\sim$ 20--25 \AA, i.e, close to the maximum value
reported in our long-term monitoring of the system (Table \ref{sp}). 
Thus, it seems reasonable to think that the disc in \ls\ reached its
maximum size by the time that the X-ray emission began.

The onset of the X-ray activity marked the beginning of the disc-loss
phase. This result can be understood since the material from the disc
constitutes the fuel that powers the X-ray source. The disc is also
responsible for the infrared excess observed in Be stars, which is
attributed to recombination radiation from ionized hydrogen (free-free and
free-bound processes). Any major change in the properties of the disc will
translate into the infrared magnitudes. Soon after the onset of the X-ray
emission, the infrared magnitudes began to decrease and by the time
H$\beta$ line went back into absorption $J$, $H$ and $K$ had decreased by
more than 0.8 magnitudes. This relationship between X-ray and infrared
emission also demonstrates that the compact object exerts a notable
influence on the fate of the Be star's disc.

Overall, the entire episode of formation and dissipation of the disc,
considering it as the time elapsed between two consecutive absorption line
phases, took $\simless$ 4.3 years (Fig \ref{beta}). It is interesting to
note that the H$\alpha$ line was showing little emission
(EW($\alpha$)=--4.6 \AA) back in April 18, 1992 (Motch et al. 1997), that
is $\sim 4$ years before the first measurement reported here. That state
must correspond with  the first stages of a previous active state, and
implies that the 4.3 yr period is repeatable. In other words, the
formation and loss of the disc represents a cyclic event in \ls.

The X-ray periodic flaring behaviour is common among Be/X-ray binaries and
the flares, which normally peak at around periastron, are referred to as
type I outbursts. They differentiate from type II outbursts which are
major increases ($10^2-10^3)$ in X-ray flux and tend to peak at later
orbital phase.  Many of the giant outbursts are detected in the middle or
at the beginning of a sequence of normal outbursts (Bildsten et al.
1997).  The case of \ls\ is different since a series of normal outbursts
followed immediately the inactive phase. In this respect \ls\ resembles
EXO 2030 +375. This 46 d orbital period Be/X-ray binary
was in an active X-ray state, between April 1991 (when BATSE became
operative) until August 1993. It switched off for three years to reappear
in April 1996, showing periodic X-ray outbursts at 46 d intervals (Reig \&
Coe 1998a; 1998b). The similar X-ray luminosities of the flares (2-4
$\times$ 10$^{36}$ erg s$^{-1}$) stresses the resemblance of these two
systems.

While the correlation between optical and infrared emission is the
expected behaviour for a growing disc the long-term monitoring of Be/X-ray
binaries is providing new insights into the way the neutron star affects
the properties of the disc and the generation of high energy radiation.
The traditional idea of expanding discs is being replaced by  disc
dissipation/reformation cycles which control the X-ray behaviour.  Once
the disc reaches a certain size, which presumably is limited by the
continuous passages of the neutron star, the disc becomes more dense and
unstable and material is fed onto the neutron star.   Negueruela et al.
(2000) have pointed out that disc truncation by the neutron star may
indirectly cause the warping/tilting of the Be star's disc. They applied
the viscous decretion disc model (Negueruela \& Okazaki 2000) to the
Be/X-ray transient 4U 0115+63/V635 Cas to explain its X-ray and optical/IR
behaviour.  4U 0115+63/V635 Cas presented a sizable disc $\sim$ 6 months
after being  observed discless and was X-ray active only 13 months after
a  photometric low. These timescales compare very well with those   in
\ls.  

The asymmetry of the H$\beta$ line profiles in LS 992 and the variability
of its blue (V) and red (R) peaks (Fig~\ref{beta}) are reminiscent of the
so-called V/R variability, and point toward a perturbed density
distribution in the disc. The perturbation is associated with the
existence of global oscillation modes propagating though the disc (Okazaki
1991, 1997; Papaloizou et al. 1992). The V/R pattern  developed shortly
after the disc build-up began, supporting the idea that the discs are
intrinsically unstable against density perturbations.

\section{Conclusion}

We have witnessed the appearance and disappearance of the circumstellar
envelope in the Be/X ray binary \ls.  Using $UBVRI$ and $uvby\beta$
photometric observations we have derived the astrophysical parameters of 
the Be/X-ray binary \ls. The optical companion turns out to be a B0.2IVe
star showing marked optical and infrared variability on timescales of
months to years. The reversal from positive to negative and then back to
positive of the equivalent width of the H$\beta$ line, the large amplitude
of variation of both, the H$\alpha$ equivalent width ($\Delta EW(H\alpha)
\approx 20$ \AA) and the infrared magnitudes ($\Delta J \approx \Delta H
\approx \Delta K \approx 0.8$ mag) are interpreted as the formation and
dissipation of the Be star's disc. The long-term monitoring of \ls\ 
constrains the timescale for formation and dissipation of the disc to
about 4.3 years. Measurements found in the literature seem to indicate
that this episode is cyclic and basically supports the picture of
dynamically unstable discs, that is, the discs go through cycles of 
dissipation/reformation which control the X-ray behaviour of the source.

\begin{acknowledgements}

P. Reig acknowledges partial support via the European Union Training and 
Mobility of Researchers Network Grant ERBFMRX/CT98/0195. IN was supported
by an ESA external fellowship. This work made use of data provided by the
{\em RXTE} ASM team at MIT and GSFC. We thank the referee, Dr. Berghoefer, for 
his useful comments.

\end{acknowledgements}


\begin{thebibliography}{99}

\bibitem{} Balona L., 1994, MNRAS 268, 11
\bibitem{} Balona L., 2000, in {\it The Be Phenomenon in Early-Type stars}, ASP
Conference Series, M.A. Smith, Henrichs H.F., Fabregat J. eds.
\bibitem{} Barnes T.G, Evans D.S., 1976, MNRAS 174, 489
\bibitem{} Bildsten L, Chakrabarty D., Chiu J., et al. 1997, ApJS 113, 367
\bibitem{} Buscombe W., 1969, MNRAS 144, 1
\bibitem{} Claret A., Gim\'enez A., 1992, A\&AS 96, 255
\bibitem{} Corbet R.H.D., Peele A.G., 2000, ApJ 530, L33
\bibitem{} Crawford D.L., 1978, AJ 83, 48
\bibitem{} Crawford D.L., Mander J., 1966, AJ 71, 114
\bibitem{} Crawford D.L., Mandwewala N., 1976, PASP 88, 917
\bibitem{} Fabregat J., Reglero V., 1990, MNRAS 247, 407
\bibitem{} Fabregat J., Torrejon J. M., 1998, A\&A 332, 643
\bibitem{} Gulati R.K., Malagnini M.L., Morossi C., 1989, A\&AS 80, 73
\bibitem{} Lamers H.J.G.L.M., 1981, ApJ 245, 593
\bibitem {} Menzies J. W., Cousins A. W. J., Banfield R. M., Laing J. D.,
1989, SAAO Circulars, 13, 1 
\bibitem{} Moon T., 1984, MNRAS 211,21
\bibitem{} Moon T., 1985, Ap\&SS 177, 261
\bibitem{} Motch C., Haberl F., Dennerl K., Pakull M., Janot-Pacheco
E., 1997 A\&A, 323, 853
\bibitem{} Napiwotzki R., Sch\"onberner D., Wenske V., 1993, A\&A 268, 653
\bibitem{} Negueruela I., Okazaki A.T., 2000, A\&A, in press
\bibitem{} Negueruela I., Okazaki A.T., Fabregat J. et al., 2000, A\&A, in press
\bibitem{} Okazaki A.T., 1991, PASJ 43, 75
\bibitem{} Okazaki A.T., 1997, A\&A 318, 548
\bibitem{} Panagia N., 1973, AJ 78, 929
\bibitem{} Papaloizou, J.C., Savonije, G. J., Henrichs, H. F, 1992, A\&A, 265, 
L45
\bibitem{} Reed C. B., 1990, AJ 100, 737
\bibitem{} Reig P, Coe M.J., 1998a, MNRAS 294, 118
\bibitem{} Reig P, Coe M.J., 1998b, MNRAS 301, 42
\bibitem{} Reig P., Roche P., 1999, MNRAS 306, 95
\bibitem{} Reig P., Fabregat J., Coe M. J. et al., 1997a, A\&A 322, 183
\bibitem{} Reig, P., Fabregat, J., Coe, M. J., 1997b, A\&A 322,193
\bibitem{} Steele I. A., Negueruela I., Clark J. S., 1999, A\&AS 137,
147.
\bibitem{} Strafella F., Pezzuto S., Corciulo G. G., Bianchini A., Vittone A. A.,
1998, ApJ 505, 299
\bibitem{} Voges W.,  Aschenbach B., Boller T et al., 1999, A\&A 349, 389
\bibitem{} Vacca W.D., Garmany C.D., Shull J.M., 1996, ApJ 460, 914
\bibitem{} Walborn N., Fitzpatrick E., 1990, PASP 102, 379
\bibitem{} Wegner W., 1994, MNRAS 270, 229

\end{thebibliography}
\end{document}